\documentclass[11pt]{article}
\usepackage{authblk}
\usepackage{rotating,amsmath,amssymb,amsfonts,amsthm,multirow,lscape,subfigure}
\usepackage{setspace}
\usepackage{epsfig,mathrsfs,natbib,color,graphics,wasysym,tikz}
\def\boxit#1{\vbox{\hrule\hbox{\vrule\kern6pt \vbox{\kern6pt#1\kern5pt}
\kern6pt\vrule}\hrule}}

\begin{document}

\title{The Application of Monte Carlo Methods for Learning Generalized Linear Model}
\author[]{Bochao Jia\thanks{Corresponding Author. E-mail: jbc409@gmail.com}}
\affil[]{
Eli Lilly and Company, Lilly Corporate Center, Indianapolis, IN 46285, U.S.A.}

\date{}

\maketitle
\renewcommand{\abstractname}{\centering\bf{Abstract}}
\begin{abstract}
Monte Carlo method is a broad class of computational algorithms that rely on repeated random sampling to obtain numerical results. They are often used in physical and mathematical problems and are most useful when it is difficult or impossible to use other mathematical methods. Basically, many statisticians have been increasingly drawn to Monte Carlo method in three distinct problem classes: optimization, numerical integration, and generating draws from a probability distribution. In this paper, we will introduce the Monte Carlo method for calculating regression coefficients in Generalized Linear Model (GLM), especially for Logistic Regression. Our main methods are Metropolis Hastings (MH) Algorithms and Stochastic Approximation in Monte Carlo Computation (SAMC). For comparison, we also get results automatically using MLE method in R software. Then we apply some Monte Carlo algorithms to a real example study and compare the efficiency of each method.  
\\
\\
\bf{Keywords: Monte Carlo method, GLM, MH-Algorithms, SAMC}
\end{abstract}
\section{Introduction}

Markov chain Monte Carlo (MCMC) methods are a class of algorithms for sampling from a probability distribution based on constructing a Markov chain that has the desired distribution as its equilibrium distribution. The Metropolis-Hastings (MH) algorithm is one of the best known of these methods, which was developed by \citet{metropolis1953equation} and subsequently generalized by  \citet{hastings1970monte}. Firstly, MH algorithm has been used in physics and was little known to statisticians until  \citet{muller1991generic} and  \citet{tierney1994markov} provided important applications using this algorithm. Then \citet{chib1996markov} gave some recent examples including relevant Markov chain theories which made more applications appeared in the recently literature. In the next section, we will provide a brief introduction to the MH-Algorithms and its work principles. We will also discuss the three different MH-Algorithms based on different sampling methods. 

As for Stochastic Approximation in Monte Carlo Computation (SAMC), it was first introduced by \citet{liang2007stochastic}. As we know, the Metropolis-Hastings (MH) algorithm and the Gibbs sampler \citep{geman1984stochastic} are prone to get trapped into local energy minima in simulations from a system for which the energy landscape is rugged. In terms of physics, the negative of the logarithmic density/mass function is called the energy function of the system. To overcome this problem, many advanced Monte Carlo algorithms have been proposed, such as parallel tempering \citep{geyer1991markov,hukushima1996exchange}, simulated tempering \citep{marinari1992simulated,geyer1995annealing}, evolutionary Monte Carlo \citep{liang2001evolutionary}, dynamic weighting \citep{wong1997dynamic}, multicanonical sampling \citep{berg1992multicanonical}, 1/k-ensemble sampling \citep{hesselbo1995monte}, the Wang-Landau algorithm \citep{wang2001efficient}, equi-energy sampler \citep{mitsutake2003replica};  stochastic approximation Monte Carlo \citep{liang2007stochastic}, among others.  In this paper, we adopt the SAMC algorithm to our problem, which has been proofed to be a sophisticated algorithm in both theory and applications. The basic idea of SAMC stems from the Wang-Landau algorithm and extends it to the continuum systems and it is designed to simulate data with a complex model structure.

The rest of the paper is organized as follows. In section 2, we will introduce three different sampling methods based on Metropolis Hastings Algorithms and SAMC algorithm for parameter estimations in the logistic regression. In section 3, we will give a simple simulation on the all of the methods and compare the results with the MLE method provided in R package. In section 4, we apply the proposed method to a  Adult Intelligence data. In section 5, we conclude the paper with a brief discussion.

\section{Method}
\subsection{Metropolis Hastings Algorithms}
The working principle of Markov Chain Monte Carlo methods is quite straight forward to describe. Given a target density $f$, we build a Markov kernel $K$ with stationary distribution $f$ and then generate a Markov chain ($X^{(t)}$) using this kernel so that the limiting distribution of ($X^{(t)}$) is $f$ and integrals can be approximated according to the Ergodic Theorem. The difficulty should thus be in constructing a kernel $K$ that is associated with an arbitrary density $f$. But, quite miraculously, there exist methods for deriving such kernels that are universal in that they are theoretically valid for any density $f$.

The Metropolis Hastings algorithm is an example of those methods. Given the target density $f$, it is associated with a working conditional density $q(y|x)$ that, in practice, is easy to simulate. In addition, $q$ can be almost arbitrary in that the only theoretical requirements are that the ratio $f(y)/q(y|x)$ is known up to a constant independent of $x$ and that $q(.|x)$ has enough dispersion to lead to an exploration of the entire support of $f$. For every given $q$, we can then construct a Metropolis Hastings kernel such that $f$ is its stationary distribution. Here is the general step of the Metropolis Hastings algorithm.

\noindent
\textbf{Algorithm} \textit{(Metropolis Hastings)}
\begin{itemize}
\item[(i)] \textit{Initialize $x^{(0)} \sim q(x)$.}
\item[(ii)] \textit{ Given current value of $x^{(t)}$, generate a candidate $y \sim q(y|x^{(t)})$.}
\item[(iii)] \textit{ Calculate acceptance probability $\alpha(x^{(t)},y)=min\{\frac{f(y)q(x^{(t)}|y)}{f(x^{(t)})q(y|x^{(t)})},1\}$.} 
\item[(iv)]\textit{Take $x^{(t+1)}=y$ with probability $\alpha(x^{(t)},y)$ and $x^{(t+1)}=x^{(t)}$ otherwise.}
\item[(v)]\textit{Iterate between step (ii) and step (iv) until converge.}
\end{itemize}

In this paper, we consider the logistic regression. We want to use Metropolis Hastings algorithm to fit the parameter of the model. i.e $\beta_0$ and $\beta_1$ in the following function.
\begin{equation}
Y_i \sim Bin(1,\pi_i), \qquad log\frac{\pi_i}{1-\pi_i}=\beta_0+x_i\beta_1, i=1,..,n
\end{equation}
And the likelihood function is
\begin{eqnarray}
&&f(\textbf{y}|\beta_0,\beta_1)=\prod_{i=1}^n \left(\frac{e^{\beta_0+x_i\beta_1}}{1+e^{\beta_0+x_i\beta_1}}\right)^{y_i}\left(\frac{1}{1+e^{\beta_0+x_i\beta_1}}\right)^{1-y_i}\\
&&\hskip 2.0cm=exp\{n\bar{y}\beta_0+\beta_1\sum_{i=1}^{n}x_iy_i-log(1+e^{\beta_0+x_i\beta_1})\}
\end{eqnarray}
Consider the prior distribution of $\beta_0$ amd $\beta_1$ as the independent normal distribution
\begin{equation}
\beta_j \sim N(\mu_j,\sigma_j^2), \quad j=0,1.
\end{equation}
Then the posterior distribution is 
\begin{eqnarray*}
&&f(\beta_0,\beta_1|\textbf{y})\propto f(\textbf{y}|\beta_0,\beta_1)\pi(\beta_0,\beta_1)\\
&&\hskip 2.0cm \propto exp\{n\bar{y}\beta_0+\beta_1\sum_{i=1}^{n}[x_iy_i-log(1+e^{\beta_0+x_i\beta_1})]\\
&&\hskip 2.0cm -\frac{(\beta_0-\mu_0)^2}{2\sigma_0^2}-\frac{(\beta_1-\mu_1)^2}{2\sigma_1^2}\}
\end{eqnarray*}
And we need to get the Markov sequence of the parameters. We propose to consider three sampling methods based on the general Metropolis Hastings algorithm, which is independent sampling, dependent sampling and individual sampling 
\subsubsection{Independent sampling}
Assuming that the parameters $\beta_0$ and $\beta_1$ are independent, we can generate $\beta^{(t)}=(\beta_0^{(t)},\beta_1^{(t)})$ from the proposal distribution $N(\beta^{(t-1)},diag(\sigma_0^2,\sigma_1^2))$. Then the independent sampling can be stated as following.
\noindent
\textbf{Algorithm} \textit{(Independent sampling)}
\begin{itemize}
\item[(i)] \textit{ Initialize $\beta^{(0)}=(\beta_0^{(0)},\beta_1^{(0)})$.}
\item[(ii)] \textit{Generate $\beta'$ from the proposal distribution $N(\beta^{(t-1)},diag(\sigma_0^2,\sigma_1^2))$.}
\item[(iii)] \textit{Calculate acceptance probability $$\alpha(\beta^{(t-1)},\beta')=min\{\frac{f(\textbf{y}|\beta_0',\beta_1')q(\beta_0',\beta_1')}{f(\textbf{y}|\beta_0^{(t-1)},\beta_1^{(t-1)})q(\beta_0^{(t-1)},\beta_1^{(t-1)})},1\}$$}
\item[(iv)] \textit{ We accept $\beta^{(t)}=\beta'$ with probability $\alpha(\beta^{(t-1)},\beta')$ or $\beta^{(t)}=\beta^{(t-1)}$ otherwise.}
\item[(v)]\textit{Iterate between step (ii) and step (iv) until converge.}

\end{itemize}
\subsubsection{Dependent Sampling}
If the parameters are not independent, the convergence of stochastic sequence may be less efficient. So we consider the multinormal prior for $\beta_0$ and $\beta_1$. In this case, we use the Fisher information matrix $H(\beta)$ to build the proposal distribution.
\begin{equation}
\beta' \sim q=N(\beta,c_{\beta}^2[H(\beta)]^{-1})
\end{equation}
where $c_{\beta}$ is a regulation parameter, which can be adjust to reach the target acceptance rate. Based on the likelihood function, we get the information matrix 
\begin{equation}
H(\beta)=X^Tdiag(h_i)X+\Sigma_{\beta}^{-1}
\end{equation}
where $\Sigma_{\beta}$ is the prior corvariance matrix of $\beta$, $h_i=exp(\beta_0+\beta_1x_i)/(1+exp(\beta_0+\beta_1x_i))^2$, $X=(\mathbf{1}_n,x)$ is a $2 \times n$ matrix.

Then we give the algorithm for this method:\\
\\
\noindent
\textbf{Algorithm} \textit{(Dependent Sampling)}
\begin{itemize}
\item[(i)] \textit{ Initialize $\beta^{(0)}=(\beta_0^{(0)},\beta_1^{(0)})$.}
\item[(ii)] \textit{Calculate Fisher Information matrix $$diag(h_i)=diag(\frac{exp(\beta_0^{(t-1)}+\beta_1^{(t-1)}x_i)}{(1+exp(\beta_0^{(t-1)}+\beta_1^{(t-1)}x_i))^2})$$
$$H(\beta^{(t-1)})=X^Tdiag(h_i)X+\Sigma_{\beta^{(t-1)}}^{-1} \quad S_{\beta^{(t-1)}}=c_{\beta^{(t-1)}}^2[H(\beta^{(t-1)})]^{-1}$$}
\item[(iii)] \textit{ Generate $\beta'$ from the proposal distribution $N(\beta^{(t-1)},S_{\beta^{(t-1)}})$.}
\item[(iv)] \textit{ Calculate acceptance probability $$\alpha(\beta^{(t-1)},\beta')=min\{\frac{f(\textbf{y}|\beta_0',\beta_1')\pi(\beta_0',\beta_1')q(\beta^{(t-1)}|\beta',S_{\beta'})}{f(\textbf{y}|\beta_0^{(t-1)},\beta_1^{(t-1)})\pi(\beta_0^{(t-1)},\beta_1^{(t-1)})q(\beta'|\beta^{(t-1)},S_{\beta^{(t-1)}})},1\}$$}
\item[(v)] \textit{ We accept $\beta^{(t)}=\beta'$ with probability $\alpha(\beta^{(t-1)},\beta')$ or $\beta^{(t)}=\beta^{(t-1)}$ otherwise.}
\item[(vi)]\textit{Iterate between step (ii) and step (v) until converge.}
\end{itemize}

\subsubsection{Individual Sampling}
In the Dependent Sampling method, we need to adjust the regulation parameter $c_{\beta}$ that make to reach the target acceptance rate. In order to avoid this process, we propose another MH sampling algorithm, i.e. Individual Sampling. For each parameter, we generate it individually from the normal proposal distribution. We only give the algorithm in two parameters condition, since when the  dimension is too large and we need to calculate acceptance rate for each parameter, which is too time-consuming.
\\
\\
\noindent
\textbf{Algorithm} \textit{(Individual Sampling)}
\begin{itemize}
\item[(i)] \textit{ Initialize $\beta^{(0)}=(\beta_0^{(0)},\beta_1^{(0)})$.}
\item[(ii)] \textit{Generate $\beta_0'$ from the proposal distribution $N(\beta_0^{(t-1)},\sigma_0^2)$.}
\item[(iii)] \textit{Set $\beta'=(\beta_0',\beta_1^{(t-1)})$ and calculate acceptance rate $$\alpha_0(\beta^{(t-1)},\beta')=min\{\frac{f(\textbf{y}|\beta_0',\beta_1^{(t-1)})\pi(\beta_0',\beta_1^{(t-1)})}{f(\textbf{y}|\beta_0^{(t-1)},\beta_1^{(t-1)})\pi(\beta_0^{(t-1)},\beta_1^{(t-1)})},1\}$$} 
\item[(iv)] \textit{We accept $\beta^{(t)}=\beta'$ with probability $\alpha_0(\beta^{(t-1)},\beta')$ or $\beta^{(t)}=\beta^{(t-1)}$ otherwise.}
\item[(v)] \textit{ Generate $\beta_1'$ from the proposal distribution $N(\beta_1^{(t)},\sigma_1^2)$.}
\vskip 0.3cm
\item[(vi)] \textit{ Set $\beta'=(\beta_0^{(t)},\beta_1')$ and calculate acceptance rate $$\alpha_1(\beta^{(t)},\beta')=min\{\frac{f(\textbf{y}|\beta_0^{(t)},\beta_1')\pi(\beta_0^{(t)},\beta_1')}{f(\textbf{y}|\beta_0^{(t)},\beta_1^{(t)})\pi(\beta_0^{(t)},\beta_1^{(t)})},1\}$$} 
\item[(vii)] \textit{ We accept $\beta^{(t+1)}=\beta'$ with probability $\alpha_1(\beta^{(t)},\beta')$ or $\beta^{(t+1)}=\beta^{(t)}$ otherwise.}
\item[(viii)]\textit{Iterate between step (ii) and step (vii) until converge.}
\end{itemize}

\subsection{Stochastic Approximation Monte Carlo Computation(SAMC)} 
Aforementioned, We introduce three MH algorithms for learning the Logistics regression. In this part, we will illustrate the Stochastic Approximation Monte Carlo Computation(SAMC) to deal with this problem.

Consider the sampling distribution that $p(x)=cp_0(x)$, where c is a constant, $x$
is generated from the sample space $\chi$. We let $E_1,...,E_m$ donate $m$ disjoint regions that from a partition of $\chi$, which can be partitioned according to any function of $x$ such as the energy function $U(x)$ as follows:
$E_1=\{x: U(x) \leq u_1\},E_2=\{x: u_1 < U(x) \leq u_2\},\ldots, E_m=\{x: U(x) \geq u_m\}$. Then we let $\hat{g}_i^{(t)}$ donate the estimate of $g_i$ and $\theta_{ti}=log(\hat{g}_i^{(t)})$ and rewritten the invariant distribution $\hat{p}(x)$ in the generalized Wang-Landau(GWL) algorithm as 
\begin{equation}
p_{\theta_t}(x)\propto \sum_{i=1}^{m}\frac{\psi(x)}{e^{\theta_{ti}}}I(x \in E_i)
\end{equation}
For theoretical simplicity, we assume that $\theta_t \in \Theta$ for all t, where $\Theta$ is a compact set.
Let $\pi=(\pi_1,...,\pi_m)$ be an m-vector with $0<\pi_i<1$ and $\sum_{i=1}^m\pi_i=1$, which defines the desired sampling frequency for each of the subregions. Henceforth,$\pi$ is called the desired sampling distribution. Let ${\gamma_t}$ be a positive, nondecreasing sequence satisfying 
\begin{equation}
(a) \quad \sum_{t=1}^{\infty} \gamma_t=\infty \qquad and \qquad (b) \quad \sum_{t=1}^{\infty} \gamma_t < \infty
\end{equation}
for some $\zeta \in (1,2)$. For example, in this article we set
\begin{equation}
\gamma_t=\frac{t_0}{max(t_0,t)}. t=1,2...,
\end{equation}
for some specified value of $t_0>1$.\\
In the logistic regression model, the invariant distribution $p_{\theta_t}(\beta) \propto exp\{n\bar{y}\beta_0+\beta_1\sum_{i=1}^{n}x_iy_i-log(1+e^{\beta_0+x_i\beta_1})\}$ and the proposal distribution $q(\beta^{(t)})=N(\beta^{(t-1)},\mathbf{1}_2))$. 
With the foregoing notation, one iteration of SAMC can be described as follows:
\\
\noindent
\textbf{Algorithm} \textit{(SAMC)}
\begin{itemize}
\item[(a)] \textit{Generate a sample $\beta^{(t)}=(\beta_0^{(t)},\beta_1^{(t)})$ by a single MH update with the target distribution given by $$p_{\theta_t}(\beta)\propto exp\{n\bar{y}\beta_0+\beta_1\sum_{i=1}^{n}x_iy_i-log(1+e^{\beta_0+x_i\beta_1})\}$$}
\begin{itemize}
\item[(i)] \textit{Generate $\beta'$ according to the proposal distribution $q(\beta'|\beta^{(t)})=N(\beta^{(t)},\mathbf{1}_2))$. If $J(\beta')\notin S$, then update $S$ to $S \cup J(\beta')$, where $S$ denote the collection of indices of the subregions from which a sample has been proposed and $J(\beta')$ denote the index of the subregion of sample $\beta'$.}
\item[(ii)] \textit{Calculate the ratio$$r=e^{\theta_{tJ(\beta^{(t)})}-\theta_{tJ(\beta')}}\frac{\psi(\beta')q(\beta^{(t)}|\beta')}{\psi(\beta^{(t)})q(\beta'|\beta^{(t)})}$$}
\item[(iii)] \textit{Accept the proposal with probability $min(r,1)$. If accepted, set $\beta^{(t+1)}=\beta'$, otherwise, set $\beta^{(t+1)}=\beta^{(t)}$.}
\end{itemize}
\vskip 0.3cm
\item[(b)] \textit{For all $i \in S$, update $\theta_t$ to $\theta_{t+1}$ as follows: $$\theta_{t+1,i}=\theta_{t,i}+\gamma_{t+1}(e_{t+1,i}-\pi_i)$$
where $e_{t+1,i}=1$ if $\beta^{(t+1)} \in E_i$, and 0 otherwise. }
\end{itemize}

\section{Simulation Study}
In this part, we generate simple logistics regression data and use the Monte Carlo method to fit the model. 
\begin{equation}
log\frac{\pi_i}{1-\pi_i}=\beta_0+x_i\beta_1, i=1,..,n
\end{equation}
Based on the model and given the value of $\beta_0$ and $\beta_1$, we generate 1000 samples of $x_i$.
\begin{equation}
x_i \sim N(1,1)  \quad \pi_i=\frac{exp(\beta_0+x_i\beta_1)}{1+exp(\beta_0+x_i\beta_1)} \quad i=1,...n
\end{equation}
Then $y_i$ is generated from binomial $\pi_i$ distribution. We do iterations for 1000 times and calculate the mean and variance of the $(\hat{\beta}_0,\hat{\beta}_1)$ in different methods.
\begin{table}[h]
\centering
\tabcolsep=3pt\fontsize{10}{14}
\caption{Comparison of parameter estimations and variances for 5 pairs of $(\beta_0,\beta_1)$.}
\vspace{0.5cm}
\label{tab:LPer}
\begin{tabular}{ccccccc}
  \hline
  \multicolumn{2}{c}{$(\beta_{0}, \beta_{1})$} & (0.1,0.2) & (0.6,0.3) & (1,-3) & (2,0.4) & (-3,2) \\ \hline
  Independent  &mean& (0.10,0.17) & (0.56,0.27) & (1.22,-3.01) & (1.78,0.35) & (-3.20,1.92)  \\ \cline{3-7}
  &variance&0.09,0.07 & 0.11,0.08 & 0.15,0.23 & 0.13,0.10 & 0.25,0.22  \\ \hline
  Dependent  &mean& (0.10,0.18) & (0.59,0.25) & (1.12,-2.87) & (2.11,0.38) & (-3.13,2.05) \\ \cline{3-7}
  &variance&0.09,0.07 & 0.09,0.08 & 0.15,0.23 & 0.13,0.10 & 0.25,0.22  \\ \hline
  Individual  & mean&(0.10,0.17) & (0.57,0.26) & (1.12,-2.88) & (2.13,0.39) & (-2.98,2.08) \\ \cline{3-7}
&variance&0.09,0.06 & 0.11,0.08 & 0.15,0.23 & 0.13,0.10 & 0.25,0.22  \\ \hline
  SAMC &mean& (0.09,0.19) & (0.62,0.27) & (1.09,-2.99) & (2.11,0.43) & (-3.19,2.01) \\\cline{3-7}
&variance&0.06,0.05 & 0.11,0.07 & 0.11,0.11 & 0.09,0.10 & 0.24,0.20  \\ \hline
  MLE & mean& (0.10,0.19) & (0.59,0.3) & (1.01,-2.97) & (2.01,0.40) & (-3.01,2.01)\\\cline{3-7}
 &variance&0.09,0.07 & 0.21,0.16 & 0.19,0.27 & 0.24,0.22 & 0.16,0.24  \\ \hline
\end{tabular}
\end{table}

Based on the Table \ref{tab:LPer}, we find that both MH algorithms and SAMC have good performance on calculating the parameters in logistics regression while SAMC has smallest variance which indicate a better convergence rate and robustness. Since the independence of $\beta_0$ and $\beta_1$, also the $Cov(\beta_0,\beta_1)$ is close to 0, all the three MH algorithms preform similarly. As for SAMC algorithm, we have to partition the sample space based on the energy function. The energy regions can set up the initial values spanned in the full model space and therefore can have faster convergence rate and lower variance than other MH algorithm. Finally, we use the MLE method,i.e the \emph{'glm'} function in R to estimate the parameters for comparison. It shows that SAMC algorithm even achieve better performance than the MLE method in terms of the variance of the estimators. Therefore, both MH and SAMC algorithms are eligible to obtain the consistent estimators in the logistic regression but even for this simple problem, SAMC still can enjoy better performance than others.

\section{Real Example}
In this part, we apply Monte Carlo methods into a real example data set. Our dataset is obtianed from the Wechsler Adult Intelligence Scale (WAIS), which is a test designed to measure intelligence in adults and older adolescents and it is currently in its fourth edition (WAIS-IV). The original WAIS was published in February 1955 by David Wechsler, as a revision of the Wechsler-Bellevue Intelligence Scale that had been released in 1939. The propose of this test is to find the relationship between intelligence and Senile Dementia among older people. In our score scale, people (with and without Senile Dementia) were asked to do some tasks individually. When all tasks completed, they will get a score, ranging from 0-20. Therefore, the response $y$ is a binary response that reflect whether people have Senile Dementia. The WAIS score is the explanation variables. Based on data, we need to build the logistic model and estimate the parameters. 

\begin{figure}[htbp]
\label{fig1}
\begin{center}
\includegraphics[scale=0.9]{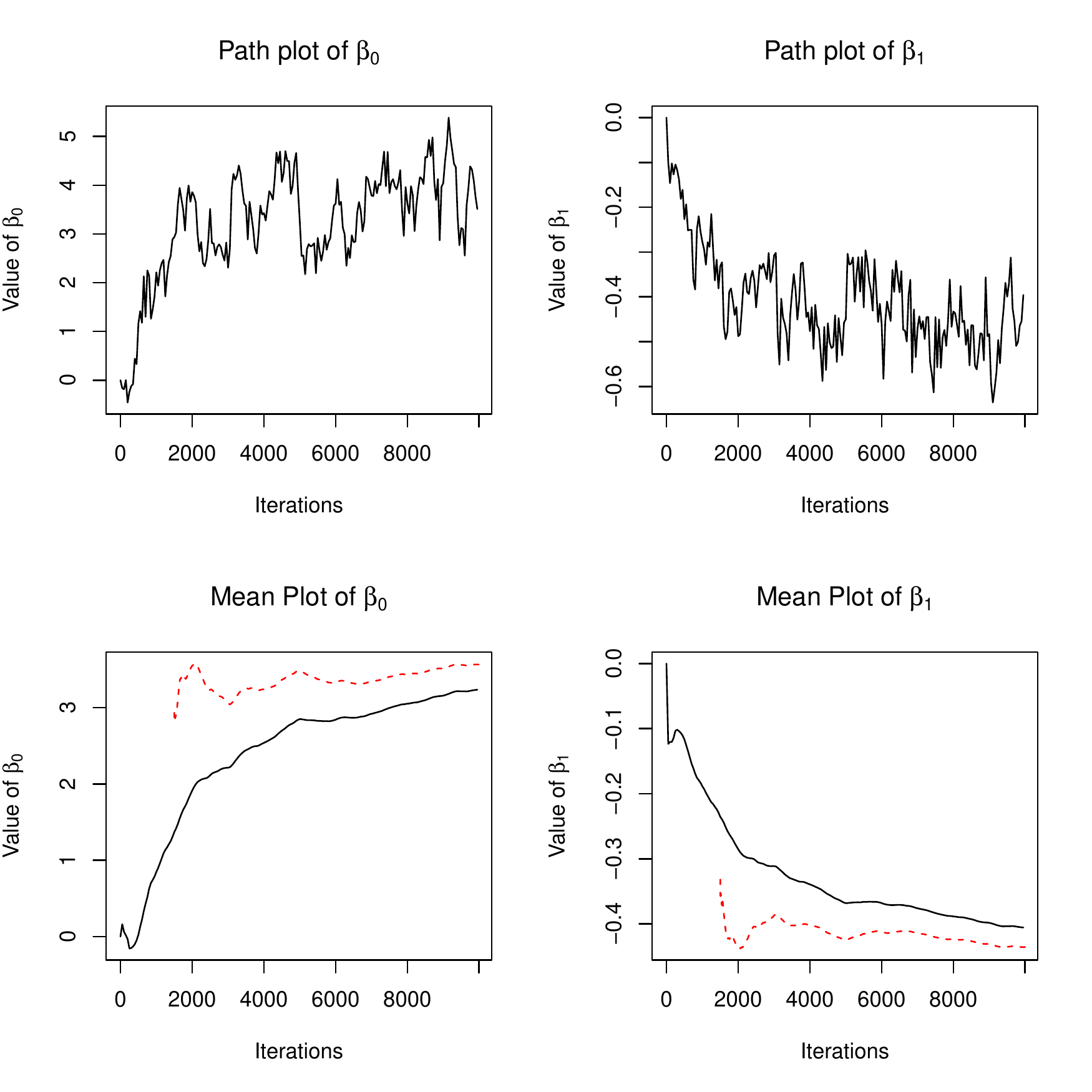}
\vskip -0cm
\caption{{The path and mean plots of $\beta_0$,$\beta_1$ for MH Independent Sampling algorithm. The upper two plots are the trace plot of estimators of $\beta_0$ and $\beta_1$ at each iteration. The lower two plots are the cumulative mean of $\beta_0$ and $\beta_1$, respectively. The solid line calculates the cumulative mean starting at iteration 1, while the dash lines starts at step 1500. }}
\label{example1}
\end{center}
\end{figure}

\begin{figure}[htbp]
\label{fig2}
\begin{center}
\includegraphics[scale=0.9]{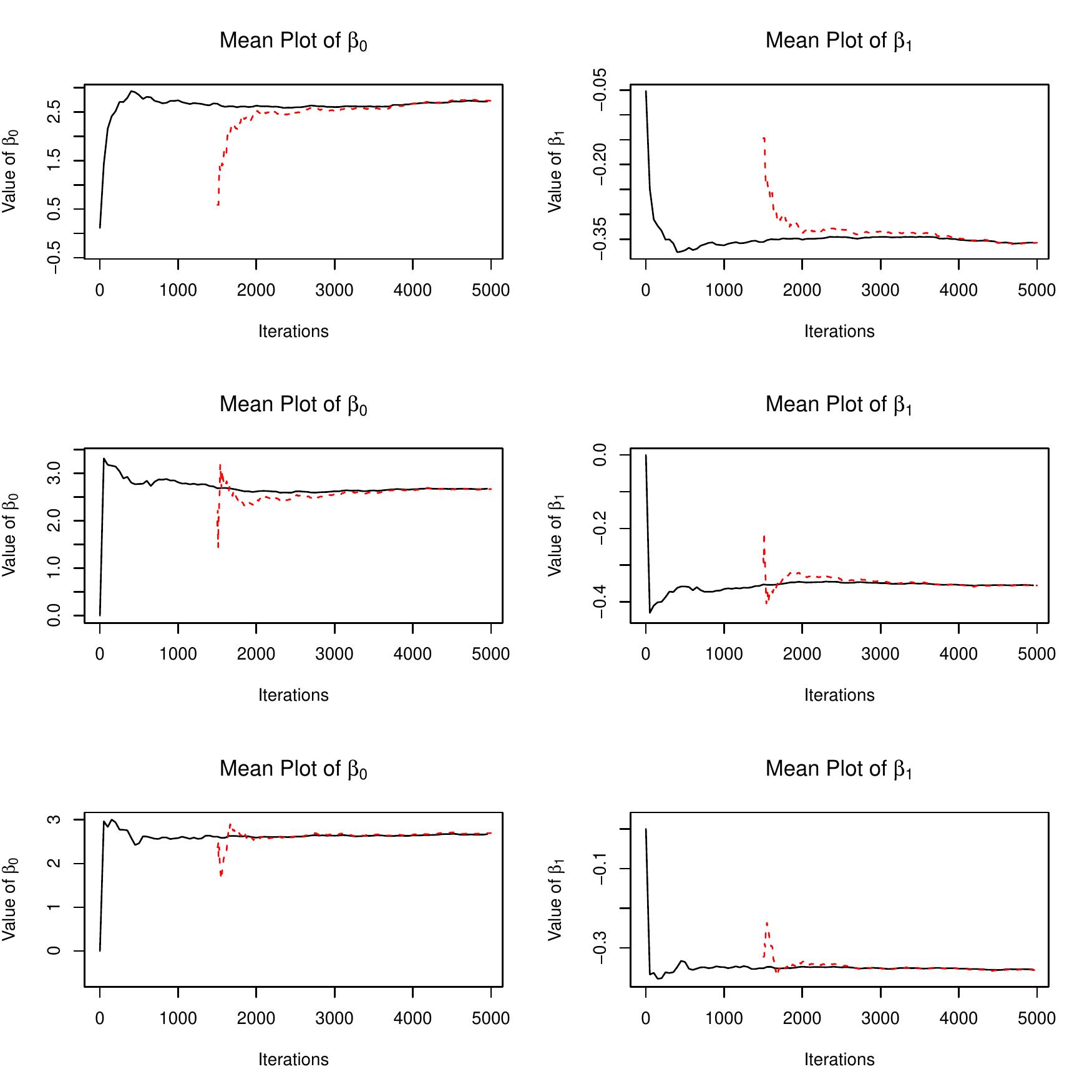}
\vskip -0cm
\caption{{Comparison the cumulative mean plot of $\beta_0$ and $\beta_1$ for multiple Monte Carlo algorithms. The upper two panels denote the MH Dependent Sampling method; the middle two panels denote the MH individual sampling method; the lower two panels denote the SAMC method. The solid line calculates the cumulative mean starting at iteration 1, while the dash lines starts at step 1500. }}
\label{example}
\end{center}
\end{figure}

Firstly, we apply the MH Independent Sampling Algorithm. We set the length of chain equals to 10000 and get the plot of parameters of $\beta_0$, $\beta_1$. As showed in the lower two figures of Figure \ref{example1}, the two lines does not coincide, which indicate that independent sampling algorithm does not converge. The independent sampling algorithm assumes that parameters are independent, however, it cannot be easily satisfied in the real application. We calculate the covariance of $\beta_0$ and $\beta_1$, which is $Cov(\beta_0,\beta_1)=-0.93$. Therefore, The independent sampling method cannot be applied here. Then we use another two MH algorithms, i.e dependent sampling and individual sampling as well as SAMC algorithm. We set the length of chain equals to 5000 and plot the cumulative mean curves for each method. As showed in Figure \ref{example}, all the methods converge when iteration grows while the SAMC converges faster than other, i.e, the two curves for SAMC coincide at around iteration 2000 while others need more iterations. To access the accuracy of estimations of $\beta_0$ and $\beta_1$, we use the results from MLE method as benchmark as compare them with the ones from the Monte Carlo methods. As showed in Table \ref{tab:LPer2}, the result from SAMC are more close to the MLE and also it has the smallest variance among all other methods. 
According to the result, $\beta_1$ is significantly less than 0, which means people who get less score have higher risk of Senile Dementia.

\begin{table}
\centering
\tabcolsep=3pt\fontsize{10}{14}
\caption{Results of both value and variance for each method of $(\beta_0,\beta_1)$}
\label{tab:LPer2}
\vspace{0.5cm}
\begin{tabular}{cccccc}
\hline 
  $(\beta_{0}, \beta_{1})$ & Independent & Dependent  & Individual  & SAMC & MLE \\ \hline
  mean& (2.68,-0.34) & (2.61,-0.34) & (2.62,-0.35) & (2.57,-0.34) &(2.40,-0.32) \\ \hline
  variance& 0.19,0.18 & 1.27,0.24 & 1.23,0.10 & 1.13,0.09  & 1.19,0.11 \\ \hline
\end{tabular}
\end{table}

\section{Discussion}
In this paper, we have proposed three MH algorithms and also SAMC to estimate the parameters of logistic regression. According to the simulation study, when the parameters are uncorrelated, each method have good performance. When the parameters are highly correlated, Independent Sampling method is less efficient and even cannot converge. The results showed that SAMC algorithms can be used in both scenarios and always outperform others. Compared with other MH algorithm, SAMC need less iterations to converge and have smaller variance of estimators even compared with MLE method. 

One interesting further work is to apply the Monte Carlo simulations to high-dimensional big data problem. The challenge comes from two aspects. First, the data contains complex structure and therefore, the acceptance rate for most MCMC algorithm can be slow and therefore can be less efficient. Although SAMC has been proved to be a good method for the complex data and can overcome some local trap problem, it still need to improve to the high-dimensional case where the likelihood function may not be intractable. Secondly, the big data have catastrophically large volumes and most single chain Monte Carlo simulations cannot applied. At each iteration, it should go through all the data which need large memory space for the computer machine and long time to generate the data at each iteration. For the big volume of data, we should consider parallel computing strategies, i.e, generate multiple Monte Carlo chains in parallel and then integrate chains to get the single estimator (Liang and Wu, 2013). Also, the high speed graphics processing unit (GPU) can be helpful for accelerating the speed of many MCMC algorithms. 

In conclusion, we introduced several Monte Carlo simulation methods to estimate parameters of generalized linear model, i.e logistic regression. This problem can also be achieved by the MLE method which might be easier to calculate. However, our Monte Carlo algorithm can obtain a chain of estimations instead of a single value and therefore can provide more information about the process of reaching the true values of parameters and can be useful for further analysis. 

\section{Acknowledgement}
The authors would like to thank the referee and the Editor for careful reading and comments which greatly improved the paper.

\bibliographystyle{asa}
\bibliography{bibdatabase}

\end{document}